\documentclass[twocolumn,
nofootinbib,
amsmath,amssymb]{revtex4-1}

\usepackage{hyperref}
\usepackage{dsfont}
\usepackage{overpic}

\newcommand{\ov}{\overline}
\newcommand{\vp}{\varphi}
\newcommand{\ccV}{\mathcal{V}}
\renewcommand{\Im}{\operatorname{Im}}
\renewcommand{\Re}{\operatorname{Re}}

\begin{document}
\title{D7-Brane Chaotic Inflation}

\author{Arthur Hebecker}
\email{A.Hebecker@ThPhys.Uni-Heidelberg.de}
\author{Sebastian C.\ Kraus}
\email{S.Kraus@ThPhys.Uni-Heidelberg.de}
\author{Lukas T.\ Witkowski}
\email{L.Witkowski@ThPhys.Uni-Heidelberg.de}
\affiliation{Institute for Theoretical Physics, Heidelberg University, Heidelberg, Germany}

\begin{abstract}
We analyze string-theoretic large-field inflation in the regime of spontaneously-broken supergravity with conventional moduli stabilization by fluxes and non-perturbative effects. The main ingredient is a shift-symmetric K\"ahler potential, supplemented by flux-induced shift symmetry breaking in the superpotential. The central technical observation is that all these features are present for D7-brane position moduli in Type IIB orientifolds, allowing for a realization of the axion monodromy proposal in a controlled string theory compactification. On the one hand, in the large complex structure regime the D7-brane position moduli inherit a shift symmetry from their mirror-dual Type IIA Wilson lines. On the other hand, the Type IIB flux superpotential generically breaks this shift symmetry and allows, by appealing to the large flux discretuum, to tune the relevant coefficients to be small. The shift-symmetric direction in D7-brane moduli space can then play the role of the inflaton: While the D7-brane circles a certain trajectory on the Calabi-Yau many times, the corresponding $F$-term energy density grows only very slowly, thanks to the above-mentioned tuning of the flux. Thus, the large-field inflationary trajectory can be realized in a regime where K\"ahler, complex structure and other brane moduli are stabilized in a conventional manner, as we demonstrate using the example of the Large Volume Scenario.

\end{abstract}

\date{14 April 2014}

\maketitle 

\section{Introduction}
The standard theory of cosmological evolution involves a period of primordial inflation which, in its simplest realization, is driven by the potential energy density of a slowly rolling scalar field, the inflaton $\vp$. This theory of slow-roll inflation is sensitive to higher-dimensional operators, thereby probing its UV completion.
Consequently, any such inflationary model needs to be implemented in a UV-complete theory of quantum gravity, such as string theory.

Models of slow-roll inflation can be classified according to the distance the inflaton rolls during inflation and are either of the large-field type, $\Delta \vp > M_p$, or of the small-field type, $\Delta \vp <M_p$. While there has been much progress in constructing small-field models in string theory (for a review see \cite{Burgess:2013sla,Baumann:2014nda}), realizing large-field models is notoriously difficult. In field theory, the latter are well studied, the prime candidate being chaotic inflation \cite{Linde:1983gd}. Crucially, in any viable representative of this class of models one needs to control all higher-dimensional operators. This is commonly done by imposing a shift symmetry for the inflaton. This symmetry is broken, e.g.\ by a term $\sim m^2 \vp^2$, with $m \ll 1 $ in units of the reduced Planck mass. The shift symmetry is restored in the limit $m\to 0$ and thus the model is technically natural in field theory.

In string theory, however, typical inflaton candidates like D-brane positions \cite{Kachru:2003sx,Dasgupta:2002ew}, Wilson lines \cite{Avgoustidis:2006zp}, and axions generically have a field range which is limited to sub-planckian values. The same is true for K\"ahler moduli \cite{Conlon:2005jm}, except where the inflaton is identified with a breathing mode of the compact space \cite{Cicoli:2008gp}. Overall, realizing large-field models in a UV-complete theory of quantum gravity is challenging.

Clearly, there are several proposed ways how one can, despite of the limited field range, construct scenarios in string theory which are effectively of the large-field type. For example, one may consider a large number of axions as in N-flation \cite{Kim:2004rp,Dimopoulos:2005ac,Kallosh:2007cc,Grimm:2007hs,Conlon:2012tz} or similar proposals \cite{Ashoorioon:2009wa, Ashoorioon:2011ki}. However, a recent analysis of an embedding of N-flation in Type IIB string theory shows that the number of axions $N$ has to be as large as $10^5$ \cite{Cicoli:2014sva}. It is questionable if such a large number can be attained.
A different interesting proposal is the use of monodromy to break the periodicity and enlarge the field space of an axion \cite{Silverstein:2008sg,McAllister:2008hb,Berg:2009tg,Palti:2014kza}, a mechanism also analyzed in field theory \cite{Kaloper:2008fb,Kaloper:2011jz,Dubovsky:2011tu,Lawrence:2012ua}. These models are plagued by control issues: In the original proposal it is a pair of NS5 and anti-NS5 branes which needs to be embedded in the compact space (see, however, \cite{Palti:2014kza}). Thus, supersymmetry is broken at the string scale and it is unclear whether the description in terms of an effective supersymmetric 4d action with the anti-branes treated as probes is valid \cite{Conlon:2011qp}.

In this letter we propose a novel way to realize large-field inflaton in string theory, using the position modulus of a D7-brane as the inflaton. Our model features the appealing mechanisms of a shift symmetry and a monodromy. Thus, in spirit it is similar to the proposals of \cite{Silverstein:2008sg,McAllister:2008hb,Berg:2009tg,Palti:2014kza}, however, with one major advantage: The model does not suffer from the control issues described above, i.e.\ it allows for a description in terms of an effective supergravity lagrangian. Furthermore, a rather explicit discussion of moduli stabilization e.g.\ in the Large Volume Scenario \cite{Balasubramanian:2005zx} is possible.

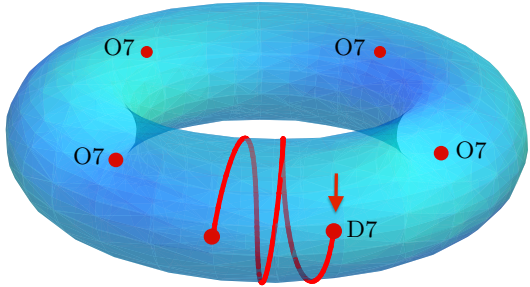
\begin{figure}
\centering
 \begin{overpic}[width=0.45\textwidth,tics=10]{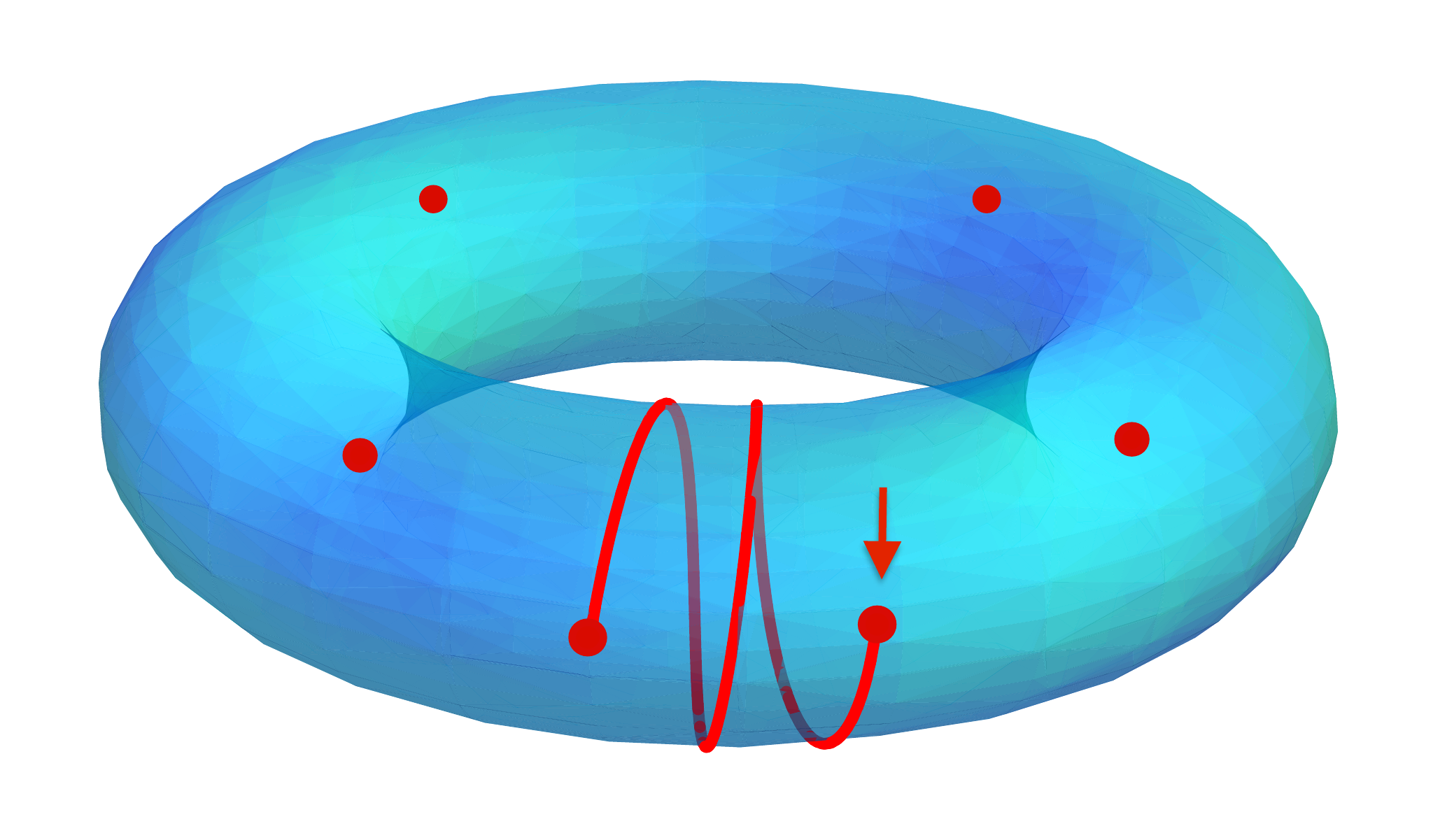}
 \put (18,26) {O7} \put (23,44) {O7}\put (61,44) {O7}\put (81,27) {O7}\put (63,14) {D7}
\end{overpic}
\caption{Illustration of the D7-brane position modulus parameter space. Inflation occurs when the D7-brane moves along a one-cycle in the parameter space, which need not necessarily be non-trivial in homology.}\label{fig:Riemann}
\end{figure}

The basic ingredients of our proposal of large field inflation with D7-branes are the following: First, we recall that the K\"ahler potential for a D7-brane modulus features a shift symmetry in the vicinity of the large complex structure point. This structure arises as the mirror dual version of the shift symmetry enjoyed by a Wilson line on a D6-brane in Type IIA string theory at large volume \cite{Kerstan:2011dy,Grimm:2011dx,Hebecker:2012qp,Hebecker:2013lha}. Disk-instantons will break the shift symmetry \cite{Kachru:2000ih}, but these effects are exponentially suppressed by the volume of the disk on the IIA side or, rather, by a complex structure modulus in the Type IIB description. The shift symmetry is crucial to avoid the supergravity $\eta$-problem \cite{Kachru:2003sx}, a mechanism equally important in the small-field cousins \cite{Hebecker:2011hk,Hebecker:2012aw,wip} of the model proposed here. Second, in the absence of fluxes the D7-brane modulus parametrizes a Riemann surface which generically has 
one-cycles, such that the field space of the modulus is periodic.\footnote{Immediateley after this work appeared the possibility of realising an inflation potential on Riemann surfaces was proposed in \cite{Harigaya:2014eta}.} In fact, all we need is a closed trajectory along the shift-symmetric direction in the D7-brane position moduli space. Fluxes will lead to an appearance of the brane modulus in the 
superpotential, such that the periodicity will be broken and a monodromy arises.\footnote{Inflation using a monodromy in the field space of a D3-brane was analyzed in \cite{Shlaer:2012by}. However, it is acknowledged in that paper that, since the proposal relies on the existence of non-trivial one-cycles in the compact space, much of the recent progress regarding moduli stabilization is not applicable in that model.} Inflation occurs along the shift-symmetric direction in the D7 moduli space. The situation is illustrated in figure~\ref{fig:Riemann}.

Displacing the D7-brane from its minimum leads to $F$-terms in the effective action which generically destabilize the potential, i.e.\ they lead to a runaway direction in the K\"ahler moduli space. Therefore, in order to ensure stability of the system during inflation, we have to tune the coefficients of the brane-modulus-dependent terms in the superpotential to small values. This can be viewed as a tuning of complex structure moduli by a suitable choice of fluxes. We assume that the landscape will provide a model with this feature and will not discuss this tuning in any detail in this letter. Rather, given the very limited understanding of large field inflation in string theory, we think it is important to demonstrate that such models can be realized in principle in a controlled string-derived supergravity framework. As a result of working in Type IIB string theory, K\"ahler moduli stabilization can be analyzed very explicitly in our model, e.g.\ in the Large Volume Scenario, and gives non-trivial constraints on the size of the overall 
volume of the compact space and the coefficients of the brane moduli in the superpotential.

An additional motivation for studying large-field inflation in string theory comes from the recent measurement of B-mode polarization \cite{Ade:2014xna} by the BICEP2 collaboration. The measured spectrum was fit in this reference to a spectrum from primordial gravitational waves, generated during an epoch of inflation. The corresponding amplitude of the tensor mode perturbations can be quantified in terms of the tensor-to-scalar ratio, which was quoted as $r = 0.2_{-0.05}^{+0.07}$. Such a large value for $r$ forces the inflaton $\vp$ to traverse a super-planckian field range during inflation \cite{Lyth:1996im,Boubekeur:2005zm}.\footnote{See, however, \cite{Avgoustidis:2008zu, Choudhury:2013iaa, Choudhury:2014kma, Antusch:2014cpa}.} Though the measurement and its attribution to primordial gravitational waves should clearly be confirmed independently, it certainly encourages our analysis of embedding a large-field model of inflation in string theory.

Related results \cite{Marchesano:2014mla, Blumenhagen:2014gta, Grimm:2014vva, Ibanez:2014kia} appeared immediately before and after this work.

\section{Ingredients}
The low energy effective description of our model is in terms of a supergravity lagrangian which is build from a K\"ahler and superpotential. Let us discuss these two quantities in more detail for our model.

\subsection{Shift-Symmetric K\"ahler Potential}
The K\"ahler potential for a D7-brane deformation modulus is given by $K \supset -\ln\left(-i(S-\ov S) - k_{\text{D7}}(u,\ov u; c, \ov c)\right)$ \cite{Jockers:2004yj,Jockers:2005zy}. Here, $S = C_0 + i/g_s$ is the axio-dilaton and $u$ denotes complex structure moduli. This K\"ahler potential arises in the weak-coupling limit from the F-theory K\"ahler potential for the fourfold complex structure moduli, given by
\begin{equation}\label{eq:GenKP}
 K = -\ln \left(\int \Omega_4\wedge \ov \Omega_4\right),
\end{equation}
where $\Omega_4$ is the holomorphic (4,0)-form of the fourfold. In the weak-coupling limit this becomes \cite{Denef:2008wq}
\begin{equation*}
 K_{g_s \to 0} = -\log\left((S - \ov S)\pi_A(u) Q^{AB}\ov\pi_B(\ov u)+ f(u,\ov u;c,\ov c)\right) + \ldots,
\end{equation*}
where $\pi_A(u)$ are the periods of the threefold, i.e.\ integrals of the holomorphic (3,0)-form over a symplectic basis of three-cycles with intersection matrix $Q^{AB}$. Furthermore, $f(u,\ov u;c,\ov c)$ is some function involving brane and complex structure moduli, but not the axio-dilaton. The above expression holds up to corrections which are suppressed at large $\Im (S) = 1/g_s$.

Under mirror symmetry \cite{Greene:1993vm}, this K\"ahler potential is identified with the K\"ahler-moduli K\"ahler potential of the mirror fourfold which is known, at large volume, to involve the volume moduli of the fourfold, but not the corresponding axions. I.e.\ the K\"ahler-moduli K\"ahler potential is shift-symmetric at large volume. Thus, via mirror symmetry we expect that \eqref{eq:GenKP} takes the shift-symmetric form
\begin{align*}
K^{\text{LCS}} = -\ln & \left( \frac{\kappa_{ijkl}}{4!}(z^i - \ov z^i)(z^j - \ov z^j)(z^k - \ov z^k)(z^l - \ov z^l) \right. \\
& \left. \ \ + \vphantom{ \frac{\kappa_{ijkl}}{4!}} \ldots\right),
\end{align*}
in the large complex structure limit \cite{Morrison:1991cd,Hosono:1994av}, which is indeed explicitly visible in the expressions derived in \cite{Honma:2013hma}. Here, $\kappa_{ijkl}$ is the self-intersection matrix of the mirror fourfold divisors, $z^i$ are the complex structure moduli of the fourfold, and the ellipses denote corrections to this shift-symmetric structure.

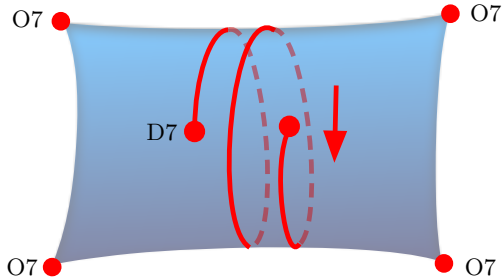
\begin{figure}
\centering
 \begin{overpic}[width=0.4\textwidth,tics=10]{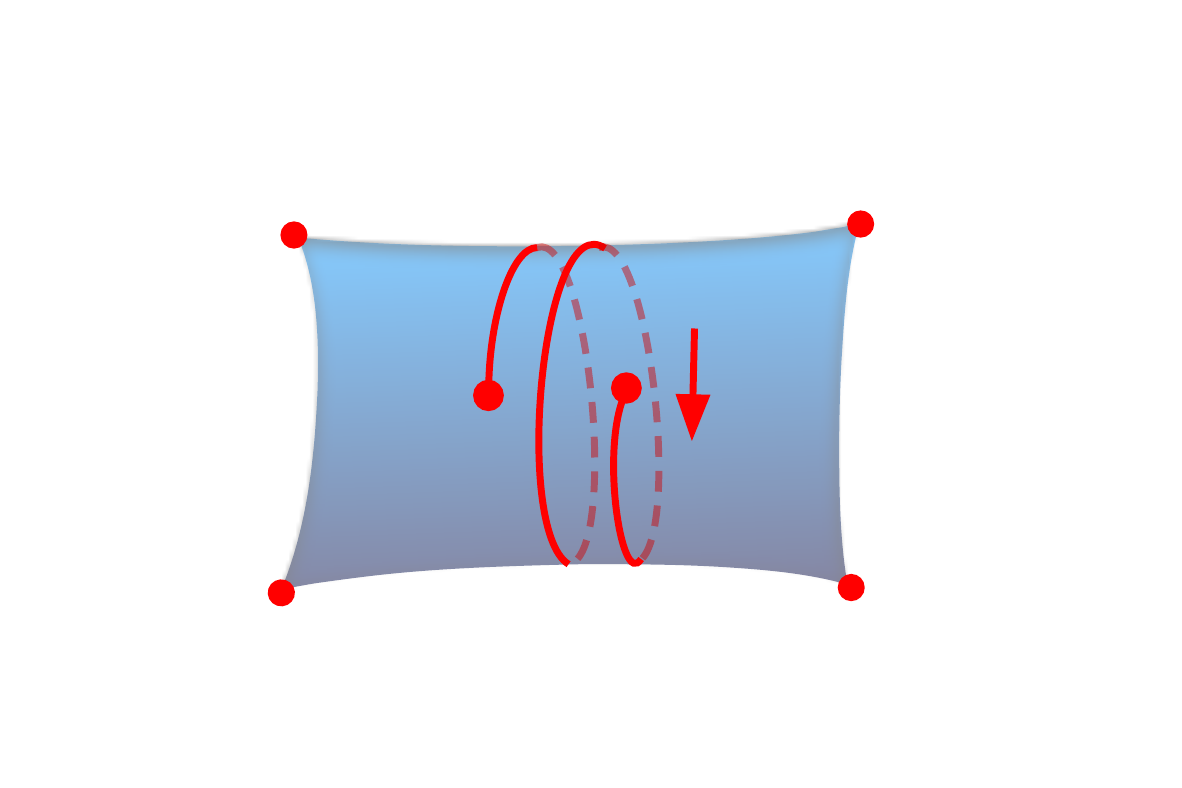}
 \put (5,57) {O7} \put (4,11) {O7}\put (90,58) {O7}\put (30,36) {D7}\put (89,11) {O7}
\end{overpic}
\caption{Illustration of the D7-brane position modulus parameter space in the example of F-theory on $K3\times K3$, which reduces to Type IIB string theory on $K3\times T^2/\mathds{Z}_2$ in the weak coupling limit.}\label{fig:K3}
\end{figure}

In the weak coupling limit, one of the $z^i$ is identified with the axio-dilaton $S$, others are identified with D7-brane position moduli $c^p$, and the rest are complex structure moduli $u^a$ of the threefold. Writing down the brane moduli dependence explicitly, we expect the structure
\begin{align*}
 K_{g_s \to 0}^{\text{LCS}} = -\ln & \left( \frac{\kappa^{(1)}_{abc}}{3!}(S - \ov S)(u^a - \ov u^a)(u^b - \ov u^b)(u^c - \ov u^c)\right.\\
 & \left. + \frac{ \kappa^{(2)}_{abpq}}{4!}(u^a - \ov u^a)(u^b - \ov u^b)(c^p - \ov c^p)(c^q - \ov c^q) \right. \\
& \left. + \vphantom{ \frac{\kappa^{(1)}_{ijkl}}{4!}} \ \ldots \ \right).
\end{align*}
Identifying one of the $c^p$ with the deformation modulus $c$ of the D7-brane with which we would like to realize inflation and integrating out all other moduli, we conjecture the following general structure for the K\"ahler potential
\begin{equation}\label{eq:ModelKahlerPot}
 K = -\ln\left(A + iB (c-\ov c) - \frac{D}{2}(c-\ov c)^2\right),
\end{equation}
where $A,B,D \in \mathds{R}$. An instructive example is F-theory on $K3\times K3$, where the K\"ahler potential is known explicitly and the shift symmetry is manifest \cite{Gorlich:2004qm,Lust:2005bd}. In the orientifold limit the model is described by Type IIB string theory compactified on $K3 \times T^2/\mathds{Z}_2$ with D7-branes and O7-planes wrapping $K3$. The parameter space of $c$ is thus $T^2/\mathds{Z}_2$, which is depicted in figure~\ref{fig:K3}. A linear term $\sim (c - \ov c)$ is not present in this example, but we think that this is a special feature of the $K3$ manifold. Indeed, we start from a generic quadratic expression in brane moduli $c^p$
and assume that all but one (called $c$) are stabilized at a high scale. Then, replacing all the heavy moduli by their vacuum values, both a quadratic and a linear term in $(c- \ov c)$ (the latter coming from mixed terms of type $(c^p-\ov{c}^p)(c -\ov c)$) arise. 
Furthermore, it is not clear to us whether terms of higher than quadratic order in $(c-\ov c)$ appear generically in the above expression. However, they would just slightly complicate the computations, but not alter our conclusions qualitatively.

Assuming that we make all the $z^i$ of the fourfold homogeneously large in the large complex structure limit, we expect the scalings $A \sim \Im(z)^4$,  $B \sim \Im(z)^3$,  $D \sim \Im(z)^2$. Here we have treated the axio-dilaton, the complex structure moduli of the threefold, and all brane coordinates except for $c$ on similar grounds. This is, of course, a very coarse approximation. As a first estimate, however, this is certainly a valid assumption.

\subsection{Superpotential}
The F-theory superpotential is given by
\begin{equation*}
 W = N^i \Pi_i(z),
\end{equation*}
where the $N^i$ are flux quantum numbers and $\Pi_i(z)$ is the period vector of the fourfold. The latter is schematically given by
\begin{equation*}
 \Pi(z) \sim (1,z^i,\kappa_{ijkl}z^i z^j,\kappa_{ijkl}z^i z^j z^k,\kappa_{ijkl}z^i z^j z^k z^l)
\end{equation*}
up to corrections. Thus, again focusing on the dependence on one brane modulus $c$, we expect the general structure
\begin{equation}\label{eq:ModelSuperpot}
 W = W_0 + \alpha c + \frac{\beta}{2}c^2 + \ldots.
\end{equation}
As an example let us once more mention the compactification of F-theory on $K3\times K3$, where precisely this structure arises. Again, we have arguments why we expect such a structure to arise also more generally. But they go beyond the scope of this letter. Here we simply assume that \eqref{eq:ModelSuperpot} is the generic structure of the superpotential.

Furthermore, as outlined in the introduction, we need to tune $|\alpha|$ and $|\beta|$ to small values in order for the induced $F$-terms to be small enough not to interfere with moduli stabilization during inflation. The merit of our model is that it indeed admits a rather explicit discussion of moduli stabilization and therefore, non-trivial constraints on $\alpha$ and $\beta$ are obtained and reported in the subsequent sections.

\section{The Model}
Given the K\"ahler potential \eqref{eq:ModelKahlerPot}, supplemented by the K\"ahler moduli part, i.e.
\begin{equation*}
 K = -2 \ln \ccV - \ln\left(A + iB (c-\ov c) - \frac{D}{2}(c-\ov c)^2\right),
\end{equation*}
and the superpotential \eqref{eq:ModelSuperpot}, supplemented by instanton corrections on small blow-up cycles
\begin{equation*}
 W = W_0 + \alpha c + \frac{\beta}{2}c^2 + e^{-2\pi T_s},
\end{equation*}
we can now write down the $F$-term potential:
\begin{equation}\label{eq:FTermPot}
 V_F = e^K \left(K^{T_\gamma \ov T_\delta}D_{T_\gamma}W \ov{D_{T_\delta}W} - 3|W|^2 + K^{c\ov c} |D_c W|^2\right).
\end{equation}
Here, the $T_\gamma$ are complexified K\"ahler moduli whose real part measures the size of a four-cycle of the threefold in units of the string length. Furthermore, $\ccV$ is the volume of the threefold. As usual, the complex structure moduli, the axio-dilaton and almost all brane moduli are assumed to be stabilized by their respective $F$-terms, with the exception of $c$ whose $F$-terms we included explicitly in \eqref{eq:FTermPot}. The reason for doing so is the very weak dependence of $W$ on $c$ which, due to the shift symmetry in the K\"ahler potential, leaves $\Re (c)$ unstabilized in a first approximation.

Owing to the fact that the K\"ahler metric is block-diagonal in the K\"ahler and complex structure moduli, no terms with mixed derivatives in $c$ and $T_\gamma$ appear in \eqref{eq:FTermPot}. Therefore, in the first two terms we can formally substitute $\tilde W_0  = W_0 + \alpha c + \frac{\beta}{2}c^2$ and the no-scale structure leads to a cancellation of the leading-order terms in $T_\gamma$. Thus, the third term in \eqref{eq:FTermPot} is dominant and stabilizes $c$ in a supersymmetric minimum, i.e.\ at $D_c W = 0$.

Now, K\"ahler moduli stabilization proceeds as in the plain-vanilla Large Volume Scenario \cite{Balasubramanian:2005zx}, giving rise to an AdS minimum which scales as $\sim -|\tilde W_0|^2/\ccV^3$. This minimum is then uplifted to a Minkowski minimum via one of the various proposed uplifting mechanisms. We are now interested in the $c$-dependence of the resulting terms, as inflation occurs along $\Re(c)$. It is clear that the terms from $-|\tilde W_0|^2/\ccV^3$ are subleading in the inverse overall volume with respect to the terms from the third term in \eqref{eq:FTermPot}. Therefore, the leading order mass term for the inflaton in our model is contained in $e^K K^{c\ov c} |D_c W|^2$. In order for this mass term not to interfere with the volume stabilization we tune $|\alpha|$ and $|\beta|$ to small values. This ensures stability in the K\"ahler moduli directions along the whole inflaton trajectory.

One crucial fact for the viability of the Large Volume Scenario is the existence of the `extended no-scale' structure \cite{vonGersdorff:2005bf,Berg:2007wt,Cicoli:2007xp} which ensures that loop corrections are subleading with respect to the $\alpha'^3$-corrections \cite{Becker:2002nn} used to stabilize the overall volume. In the above references it is generally assumed that complex structure moduli, the axio-dilaton and all brane moduli are integrated out at a higher scale. More generally, the extended no-scale structure persists even if the low energy theory includes a complex scalar which does not appear in the superpotential and which has a shift-symmetric component in the K\"ahler potential, such that this component remains light. This will be demonstrated explicitly in \cite{wip}. Clearly, in our setting this structure will be broken by the explicit dependence of the superpotential on $c$. However, since the extended no-scale structure is restored in the limit of vanishing $\alpha$ and $\beta$, the 
breaking will be small in the limit of small $|\alpha|$ and $|\beta|$ and the overall picture remains consistent.

\subsection{Minimizing the Potential}
Let us analyze the stabilization of $c$ in more detail. We will work in the limit of small $|\alpha|$ and $|\beta|$ throughout. From $D_c W = 0$ we obtain the equation
\begin{equation}\label{eq:Stabilization}
 \frac{\alpha + \beta c}{W_0 + \alpha c + \frac{\beta}{2}c^2} = \frac{i B - D(c - \ov c)}{A + iB (c-\ov c) - \frac{D}{2}(c-\ov c)^2}.
\end{equation}
In the following we will write $c = x + iy$ with $x,y \in \mathds{R}$.
At $0^{\text{th}}$ order in $\alpha$ and $\beta$, the left-hand-side of this equation vanishes and $y $ is stabilized at
\begin{equation}\label{eq:yStab}
 y_0 = \frac{B}{2D}.
\end{equation}
Furthermore, we observe that the RHS of \eqref{eq:Stabilization} is purely imaginary. Requiring the real part of the LHS to vanish leads, in $1^{\text{st}}$ order in $\alpha$ and $\beta$, to
\begin{equation*}
 x_0  = \frac{\Im(\beta \ov W_0)y_0 - \Re(\alpha \ov W_0)}{\Re(\beta \ov W_0)}.
\end{equation*}
Thus, recalling the scaling of $A$, $B$, and $D$ with $\Im (z)$ we find $y_0\sim x_0 \sim \Im (z)$. These expressions will be corrected at higher order in $\alpha$ and $\beta$. However, since these coefficients have to be tuned to small values anyhow, for our purposes the above analysis is sufficient. 

\subsection{Computing the Mass}
We now compute the mass for the inflaton. As motivated above, the mass term will arise from $|D_c W|^2$. Furthermore, since $D_c W = 0 $ in the minimum, it suffices to expand this term in leading order in the variation of the real part of $c$, i.e.\ in $\delta x$. Furthermore, since stabilization enforces $K_c = W_c / W $ (cf.\ \eqref{eq:Stabilization}) and the latter scales linearly with $\alpha$ and $\beta$, displacing $x$ from its minimum simply gives
\begin{equation*}
 \delta D_c W =\delta \left( K_c W + W_c\right) \simeq \delta W_c  =  \beta \delta x
\end{equation*}
in linear order in $\alpha$ and $\beta$, leading to
\begin{equation*}
 e^K K^{c\ov c} |\beta|^2 \delta x^2 + \text{higher order in } \alpha, \beta, \delta x .
\end{equation*}
Now, $\delta x$ is related to the inflaton via canonical normalization. The kinetic term for $\delta x$ is contained in $K_{c \ov c} |\partial c|^2$. Recalling the scaling $K_{c\ov c} \sim \Im (z)^{-2}$ and $e^K \sim \Im (z)^{-4}$, we find
\begin{equation*}
 m_{\vp}^2 \sim \frac{1}{\ccV^2} \frac{1}{\Im (z)^4}\Im (z)^2 \Im (z)^2 |\beta|^2 = \frac{|\beta|^2}{\ccV^2},
\end{equation*}
where the two factors of $\Im (z)^2$ come from canonically normalizing the inflaton and from the $K^{c\ov c}$ factor in the $F$-term potential, respectively. Interestingly, $\Im (z)$ does not show up in $m_{\vp}^2$.

\section{Phenomenology}
The phenomenology of quadratic inflation is, of course, well known \cite{Linde:1983gd}. Let us briefly recall the basic statements. For a potential $V = m^2 \vp^2$ the slow-roll parameters are determined as
\begin{align*}
 \epsilon &= \frac{1}{2}\left(\frac{V'}{V}\right)^2 = \frac{2}{\vp^2} ,\\
 \eta &= \frac{V''}{V} = \frac{2}{\vp^2}.
\end{align*}
The spectral index can be expressed in terms of these two quantities as
\begin{equation*}
 n_s -1 = -6\epsilon + 2 \eta = -\frac{8}{\vp^2}.
\end{equation*}
Since this quantity is measured to be $\simeq -0.04$ \cite{Ade:2013uln}, the field displacement at the beginning of the last $\sim 60 $ $e$-folds of inflation is determined to be $\vp^2 \simeq 200$. The tensor-to-scalar ratio is thus fixed as
\begin{equation*}
 r = 16 \epsilon \simeq 0.16.
\end{equation*}
On the other hand, the measured value for the amplitude of curvature perturbations determines \cite{Ade:2013uln}
\begin{equation*}
 \sqrt{\frac{V}{2\epsilon}} = 5.1 \cdot 10^{-4},
\end{equation*}
which leads to $m \simeq 0.5 \cdot 10^{-5}$.

This can be translated into requirements on our stringy model of large field inflation. In particular,
\begin{equation}\label{eq:SizeOfMass}
 m_\vp = \frac{|\beta|}{\ccV} \stackrel{!}{=} 0.5 \cdot 10^{-5}.
\end{equation}
This is, however, not the only constraint which $m_\vp$ has to satisfy. As mentioned before, in order not to interfere with K\"ahler moduli stabilization we need to require
\begin{equation}\label{eq:StabilityRequirement}
 m_\vp^2 \vp^2 \simeq 0.5 \cdot 10^{-8} \ll \frac{|W_0|^2}{\ccV^3}
\end{equation}
along the whole inflationary trajectory.

To give a few specific numbers, let us choose $\ccV = 10^3$. This determines, via \eqref{eq:SizeOfMass}, $|\beta| = 0.5 \cdot 10^{-2}$. Then, \eqref{eq:StabilityRequirement} is satisfied for $|W_0| = 10$. But this is by no means the only possible realization: Also a choice $\ccV = 10^2$, leading to $|\beta| = 0.5 \cdot 10^{-3}$, works fine, even for $|W_0| = 1$.

\subsection{Stability during Inflation}
During inflation, the real part of $c$ traverses a large distance in field space. We should thus make sure that the stabilization of the imaginary part is not significantly affected by this field displacement. Recall that the kinetic term for $x = \Re(c)$ reads
\begin{equation*}
 K_{c\ov c} (\partial \delta x)^2 \sim \frac{(\partial \delta x)^2}{\Im (z)^2}.
\end{equation*}
At the beginning of the last $N \simeq 60$ $e$-folds of inflation, the canonically normalized inflaton $\vp$ takes the value $\vp_N \simeq 14$, giving
\begin{equation*}
 \delta x_N \sim 14 \cdot \Im (z).
\end{equation*}

Now consider the stabilization equation \eqref{eq:Stabilization}. One can easily convince oneself that, writing $y = y_0 + \delta y$, the consistency requirement $|\delta y| \ll y_0$ is satisfied as long as
\begin{equation*}
 \frac{ 14 |\beta| \Im (z)^2 }{|W_0|} \ll 1.
\end{equation*}
Choosing $\beta = 0.5 \cdot 10^{-3}$ and $|W_0|=1$, $\Im (z)$ is constrained as
\begin{equation}\label{eq:UpperZBound}
 \Im (z) < 12 .
\end{equation}
This potentially presents a conflict with the large complex structure limit. However, since the suppression of the correction terms is of exponential nature, even in view of \eqref{eq:UpperZBound} one can choose $z$ large enough in order to suppress these corrections.

\subsection{Cubic and quartic terms}
Beyond the mass term the inflaton potential will also exhibit cubic and quartic terms in $\delta x$. Expanding the potential \eqref{eq:FTermPot} in $\delta x$ about the minimum one finds
\begin{align}
\nonumber V \sim \frac{|\beta|^2 \ \  \delta x^2}{\mathcal{V}^2 \ {\Im (z)^2}} & \left\{ 1 + \mathcal{O} \left( \frac{\alpha^3}{\beta W_0^2}, \frac{\alpha^2 c_0}{ W_0^2}, \frac{\alpha \beta c_0^2}{W_0^2}, \frac{\beta^2 c_0^3}{ W_0^2} \right) \delta x \right. \\
& \left. \quad \ + \ \mathcal{O} \left(\frac{\alpha^2}{ W_0^2}, \frac{\alpha \beta c_0}{W_0^2}, \frac{\beta^2 c_0^2}{ W_0^2} \right) \delta x^2 \right\} \ . \label{ExpandPot}
\end{align}
The above is derived by first expanding \eqref{eq:FTermPot} in both $\delta x$ and $\delta y$ about $c_0=x_0 + i y_0$. Further we solve $D_c W =0$ for $y=y_0 + y_1$ to first order in $\alpha, \beta$. Equation \eqref{ExpandPot} is finally obtained by substituting $\delta y = y_1$.

Here we examine the relevance of the cubic and quartic terms at the onset of the last 60 $e$-folds of inflation at $\delta x_N \sim 14 \cdot \Im (z)$. As $|c_0| \sim \Im (z)$ we find that the parametrically most important term is 
\begin{equation}
\label{quartic}
V \supset \frac{|\beta|^2}{\mathcal{V}^2 \ {\Im (z)^2}} \ \mathcal{O} \left(\frac{\beta^2 c_0^2}{ W_0^2} \right) \delta x^4 \ .
\end{equation}
Terms involving $\alpha$ are not dangerous as we can always tune $\alpha$ independently of the phenomenological discussion above. For $|\beta| = 0.5 \cdot 10^{-3}$ and $|W_0| = 1$ we find that, if we set $\Im (z) \sim 10$, the term \eqref{quartic} becomes comparable to the mass term at the onset of the last 60 $e$-folds:
\begin{equation*}
\frac{|\beta|^2 |c_0|^2}{ |W_0|^2} \delta x_N^2 \sim 0.25 \cdot 10^{-6} \cdot 10^2 \cdot 200 \cdot 10^2 \sim 1 \ .
\end{equation*}
For larger $\delta x$ we then transition to a regime where the potential is dominated by the quartic term. However, such a large $\Im (z)$ is close to the upper bound \eqref{eq:UpperZBound}. For $\Im (z) < 10$, the quartic term is still subleading compared to the mass term at $\delta x_N$. In this case the quartic term in \eqref{ExpandPot} can be made comparable again by choosing an appropriate value for $\alpha$. Such corrections to the inflaton potential have been discussed recently in \cite{Kaloper:2014zba}.

\subsection{Corrections}
So far we have completely neglected the mirror-dual version of the Type IIA worldsheet instanton corrections to the K\"ahler potential. These are expected to give oscillatory contributions at the order
\begin{equation*}
 \sim e^{-2\pi y_0} \frac{|W_0|^2}{\ccV^2}
\end{equation*}
to the $F$-term potential. Thus, in view of \eqref{eq:yStab}, they are exponentially suppressed in the limit of large complex structure.
Furthermore, loop corrections due to the exchange of Kaluza-Klein modes between branes \cite{Berg:2005ja,Berg:2005yu,Berg:2007wt,Cicoli:2007xp} will also lead to periodic corrections, roughly at the order
\begin{equation*}
 \sim \{\alpha, \beta\}\cdot \frac{|W_0|^2}{\ccV^{8/3}}.
\end{equation*}
The induced corrections can be parametrized at leading order as
\begin{equation}\label{eq:PeriodicCorr}
 V = m^2 \vp^2 + \gamma \cos \left(\frac{\vp}{f} + \delta\right).
\end{equation}

The phenomenology of such a periodic modulation of a monomial inflaton potential, in particular their effect on the power spectrum and the bispectrum, was investigated for axion monodromy inflation (i.e.\ with a linear rather than a quadratic potential) in \cite{Flauger:2009ab} and more generally in \cite{Flauger:2010ja,Kaloper:2011jz}.
Since the axion decay constant is small, roughly bounded by $f \lesssim 1/4\pi$ (see \cite{Hebecker:2013zda} and references therein), during the initial observable $e$-folds the inflaton typically crosses more than one period of the oscillatory piece in \eqref{eq:PeriodicCorr}. Thus, if present and sufficiently large, the oscillatory features leave their imprint in the observable CMB modes. An explicit computation of the oscillatory terms in our model and a detailed analysis of the observational implications along the lines of \cite{Kaloper:2011jz} would be interesting, but is beyond the scope of this letter. In any case, the periodic modulations become small in the limit of large $\Im (z)$ and small $|\alpha|$ and $|\beta|$.

\section{Conclusions}
In this paper, we have outlined a scenario which has the potential to realize large-field inflation in Type IIB string theory, within a controlled 4d supergravity description with conventional moduli stabilization. More specifically, our inflaton is a D7-brane position modulus with shift-symmetric K\"ahler potential. This shift symmetry is inherited from the shift symmetry of a D6-brane Wilson line in the mirror-dual Type IIA model. Furthermore, since this latter shift symmetry requires large volume, we need to be at large complex structure in our Type IIB scenario. Shift-symmetry-breaking corrections to the K\"ahler potential are exponentially suppressed in the  large periods of the complex-structure and D7-brane moduli space. Hence, they are relatively easy to control.

The inflaton potential is quadratic at leading order. It is induced by the flux-superpotential which also depends on D7-brane positions. The coefficients of the relevant terms in the superpotential depend on complex structure moduli and other D7-brane positions. They can hence be tuned to be small, given a sufficiently large flux discretuum. As a result, the coefficient of the quadratic inflaton potential (i.e. the inflaton mass) can be made small. 

Clearly, going to a large VEV of the D7-brane position is impossible within the standard D7-brane moduli space, which is rather small. However, the natural periodicity of this space is broken by the flux mentioned above, such that a non-trivial monodromy arises. Thus, large-field inflation arises because a D7-brane circles a closed trajectory in its moduli space many times, thereby slowly growing a significant contribution to the $F$-term potential.

Our parametric analysis demonstrates that the above-mentioned flux-tuning allows us to prevent this contribution from destabilizing other moduli. We analyzed a concrete example based on the Large Volume Scenario, where the most dangerous destabilization direction is that of the overall volume. However, a tuning of the coefficients to about $10^{-3}$ of their natural value, combined with an overall superpotential $W_0\sim 1$ and a volume ${\cal V}\sim 10^2$, allows us to escape destabilization. 

Finally, we have also analyzed naively sub-leading (in the fine-tuned small coefficients) effects which correct the quadratic form of the potential. Very interestingly, it turns out that some of these effects can become considerable in the region of inflation relevant for the presently observed CMB perturbations. 

Many open questions had to be left for future work. They include an explicit demonstration of the flux-based tuning, a more detailed phenomenology of the inflationary potential, the combination of our D7-brane inflation scenario with other K\"ahler moduli stabilization mechanisms, and the discussion of corrections associated with the uplifting contribution. 

\vspace*{0.5cm}
\begin{acknowledgments}
We would like to thank Timo Weigand, Eran Palti, Stefan Sj\"ors, and Dominik Neuenfeld for helpful discussions. This work was supported by the Transregio TR33 ``The Dark Universe''.
\end{acknowledgments}

\bibliography{QuadraticInfl}
\end{document}